# The elastic properties of composites reinforced by a 3D Voronoi fibre network with or without missing fibres

Zhengyang Zhang[1,2], Hanxing Zhu[2*], Ru Yuan[3], Sanmin Wang[3], Yanming Wang[1], Tongxiang Fan[4], Yacine Rezgui[2], Di Zhang[4]

[1] UM-SJTU Joint Institute, Shanghai Jiao Tong University, Shanghai, 200240, China

[2] School of Engineering, Cardiff University, Cardiff CF24 3AA, UK

[3] School of Mechanical Engineering, Northwestern Polytechnical University, Xi'an 710072, China

[4] State Key Lab of Metal Matrix Composites, School of Materials Science and Engineering, Shanghai Jiao Tong University, Shanghai 200240, P.R. China

[*] e-mail: zhuh3@cardiff.ac.uk

**Abstract**

Many composite materials, both natural and fabricated, process a Voronoi like architecture or microstructure. Furthermore, the stochasticity and connectivity of Voronoi tessellation endow the composite materials constructed by this kind of structures a wide range of desired properties including high and tuneable stiffness, high strength and good manufacturability. Thus, Voronoi-based fibre network structures are regarded as promising designs of composite reinforcements, as well as powerful tools in composite mechanics simulations. In this paper, the elastic properties of composites reinforced by a 3D Voronoi fibre network are systemically investigated based on the precise control of the Voronoi cell regularity. The regularity of the reinforcement fibre networks, according to our definition, is found positively related to the Young's moduli of the composites. Interestingly, 20% percent of defects in total reinforcement fibres only causes a less than 6.5% of Young's moduli drop in Voronoi fibre network reinforced composite. The Voronoi fibre network reinforced composites also shows higher Young's moduli than those of conventional composites with discrete reinforcements. The possibility of simulating aerogels by Voronoi fibre network structures are also presented.

# 1. Introduction

Interpenetrating phase composites (IPCs) reinforced by a self-connected cellular network remain a research hotspot in various fields including aerospace, construction, biomaterial applications etc. [1–8]. Cellular structures with regular [9] or random [10] cells in these composites can be regarded as 3D tessellations, and Voronoi tessellations named after George Voronoy [11] are still the most popular and effective geometric models to simulate stochastic fibre networks. As various natural and fabricated cellular structures have geometries very similar to 2D [12] or 3D Voronoi tessellations, they are often used to represent the real structures and to capture their mechanical behaviours. For example, the recently developed $La_2Zr_2O_7$ ceramics with both low thermal conductivity and high compressive strength have hollow-grained "Voronoi foam" microstructure [13]. Voronoi-based composite structure with polygonal tablets and organic adhesive bonding is capable of closely mimicking nacre's multilayer composite structure and capturing its deformation and mechanical behaviours [14]. Voronoi cell finite element method (VCFEM) [15,16] has been developed and extended (X-VCFEM) [17] to simulate the elasticity and cracking in particle composite materials. Voronoi tessellations are also employed as the models of aerogels [18] and used to simulate the mechanical behaviour of biological tissues [19].

In the perspective of mechanical performance, composites reinforced by Voronoi fibre networks possess better stiffness and strength, property tunability, and ability to keep structural integrity with partial damage or imperfection, compared with the conventional particle-reinforced or fibre-reinforced composites. The properties of interpenetrating composites depend not only on the volume fractions and mechanical properties of the constituent materials, but also on the architectures and microstructures of the reinforcement fibre network and the matrix. The effects of fibre density and cell irregularity on the mechanical properties of Voronoi honeycombs [12] and open-cell foams [10,20] have been well investigated, but it is less known how the microstructures of the Voronoi fibre network reinforcements affect the mechanical properties of the corresponding interpenetrating composites. Furthermore, the difference between the regular lattice reinforced composites and stochastic fibre network reinforced composites in terms of mechanical performance is still unveiled. Thus, the effects of different architectures and microstructures of the reinforcement fibre network and the matrix on the mechanical properties of the composites, as well as the mechanical property changes of the composites caused by partial damage/imperfection of the reinforcement fibre network need to be investigated.

The aim of this paper is to explore the effects of microstructure of the reinforcement fibre-network and its defects on the elastic properties of interpenetrating composites. Interpenetrating composites are usually manufactured by producing the reinforcement network (e.g. an open cell foam) first, then casting a liquid state matrix material (e.g. epoxy) into the porous space of the network. The microstructure of the reinforcement random fibre network is modelled by a 3D Voronoi tessellation with a controlled degree of regularity [10,20]. Periodic representative volume elements (RVEs) are developed to model the elastic properties of the interpenetrating composites (IPCs). The elastic properties of the RVEs are predicted using finite element analysis and the periodic boundary conditions. The effects of the missing fibres, the numbers of complete cells and the degree of regularity of the reinforcement Voronoi open cell foam on the elastic properties of the interpenetrating composites (IPCs) are investigated respectively. Moreover, the influences of elastic properties of the constituent materials on those of the IPCs are also explored.

## 2. Geometric structures and computational methods
### 2.1. The reinforcement Voronoi fibre network structure

Periodic cubic representative volume elements (RVEs) with an edge length $L$ are developed to model the elastic properties of interpenetrating composites (IPCs) reinforced by a 3D random Voronoi fibre network. The periodic reinforcement random Voronoi open cell foam in a cubic space $L \times L \times L$ is constructed in the same way as given in [10,20]. For simplicity of the model, all the fibres are assumed to have the same uniform circular cross-section of diameter $d$, thus, the density (i.e. the solid volume fraction) of the reinforcement open cell foam can be controlled by the total length of the fibres and the fibre cross-sectional area. Figure 1 shows the periodic RVE models of IPCs reinforced by a fibre-network (open cell foam) with different numbers of complete cells.

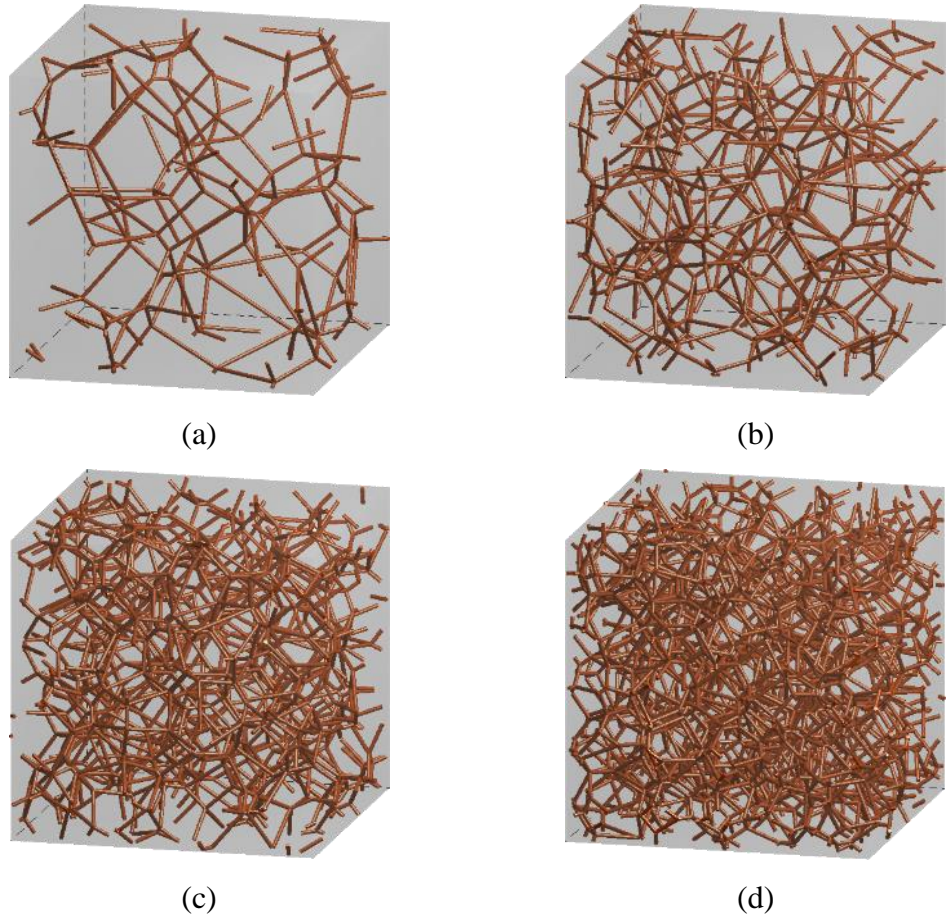

(a) (b) (c) (d)

Figure 1. 3D view of the RVEs of IPCs reinforced by Voronoi fibre networks with different numbers of complete cells: (a) $N = 16$ (b) $N = 54$ (c) $N = 128$ (d) $N = 250$.

**2.2. Degree of cell regularity**

The degree of regularity of a Voronoi open cell foam is given by [10,20]. To construct a regular open cell foam with $N$ identical body-centred cubic (BCC) cells in a space of $L \times L \times L$, the minimum distance between any two adjacent cells is $d_0$. To construct a random Voronoi open cell foam with $N$ cells in the same space, the minimum distance between any two adjacent cells is $\delta_{min}$. The degree of cell regularity of the reinforcement random open cell foam can be determined as [10,20]

$$R = \frac{\delta_{min}}{d_0} = \frac{2^{1/6} N^{1/3} \sqrt{6} \delta_{min}}{3L} \tag{1}$$

Thus, for a fully regular Voronoi tessellation with identical BCC cells, $R = 1$; while for a completely random one, $R = 0$; Detail of model construction and regularity calculation could be found in supplementary material S1.

For example, the Voronoi fibre network reinforced composite with 16 cells and different $R$s are shown in Figure 2 below.

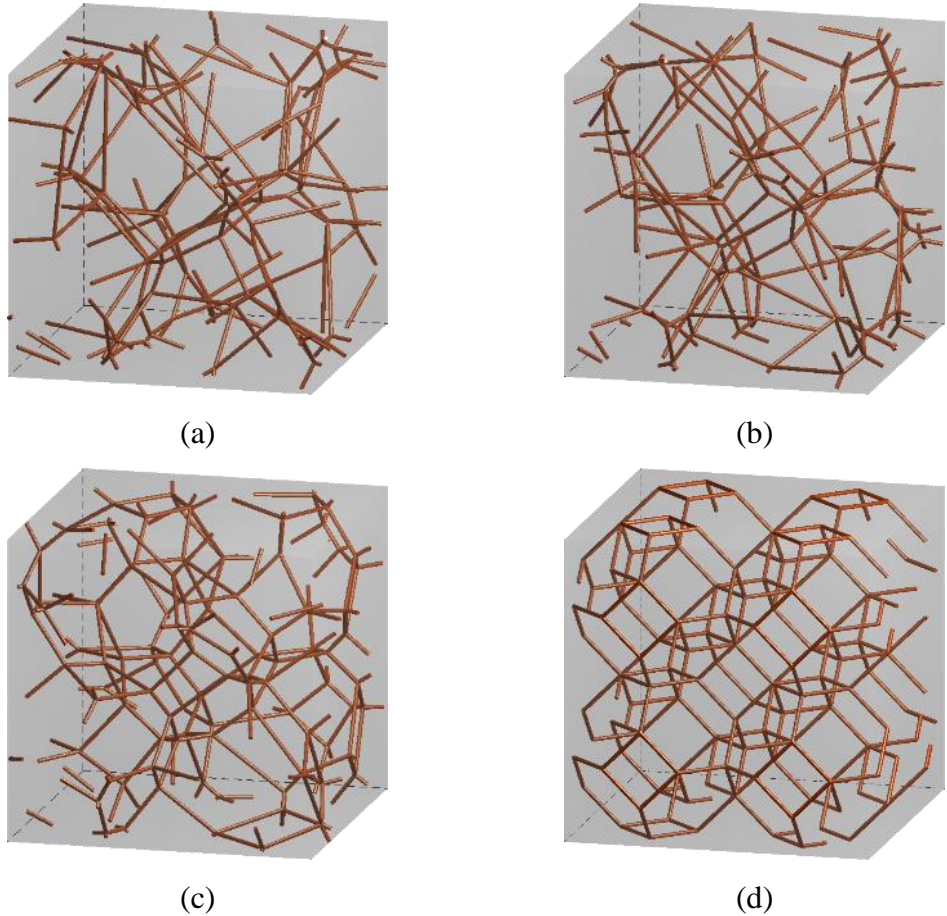

(a)          (b)

(c)          (d)

Figure 2. Illustration of cell regularity of the reinforced Voronoi fibre network with 16 cells ($N = 16$) (a) $R = 0$ (b) $R = 0.33$ (c) $R = 0.67$ (d) $R = 1$.

## 2.3. Defects by missing fibres

Defects are introduced by randomly removing a certain percentage of fibres (i.e. cell struts) from the reinforcement fibre network, representing the fact that the defects (i.e. missing fibres) exist in the manufactured/synthesized reinforcement Voronoi fibre network (i.e. open cell foam). Various numbers of fibres are removed to illustrate the effects of missing fibres on the elastic properties of the IPCs. Different degrees of defects, noted as $D = 3\%$, $5\%$, $10\%$ and $20\%$, are represented by randomly removing the corresponding numbers of fibres from the reinforcement network of the RVE as shown in Figure 3. For all the RVEs of the same structure with different numbers of missing fibres, the volume fractions of the reinforcement fibre networks in the IPCs are kept the same in further simulation by adjusting the cross-sectional area of the fibres.

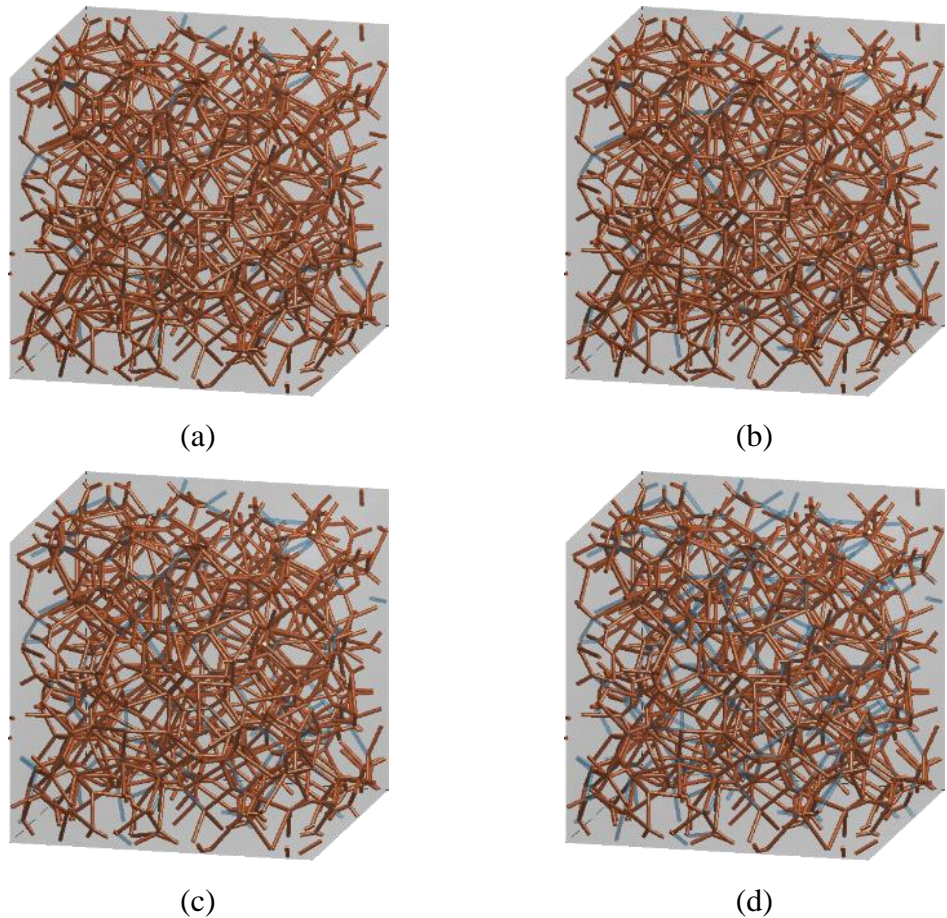

Figure 3. Illustration of fibre network defects by randomly removing fibres (blue-coloured struts) from the reinforcement Voronoi fibre network with $N = 128$ and $R = 0.5$ (a) $D = 3\%$ (b) $D = 5\%$ (c) $D = 10\%$ (d) $D = 20\%$.

## 2.4. Pre-processing and solving

All the struts in the reinforcement open cell foam are assumed to have the same circular cross-section and to be rigidly connected. 3D three-node Timoshenko beam elements (beam 189 in ANSYS [21] software) are used to represent the struts of the reinforcement Voronoi open cell foam for efficiency in simulation. This is because if using solid elements to represent all the struts of the reinforcement open cell foam, the total number of elements of a single composite RVE model could be much more than one million. Although there is no overlap between the reinforcement Voronoi fibre network and the matrix in IPCs, the difference between the elastic properties of the RVE models obtained by treating the reinforcement fibre network either as beam elements (which have overlap with the matrix) or as solid elements (without overlap with the matrix) is not significant (see Appendix, Figure A1). Figure 4 shows the periodic reinforcement 3D Voronoi fibre network (with 16 cells) in a RVE model of the IPC.

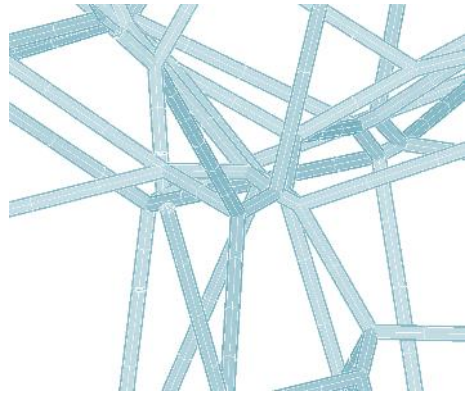

Figure 4. Enlarged 3D Voronoi fibre network reinforcement represented by beam elements.

The matrix of the RVE is assumed to be isotropic in elastic properties, which is partitioned into 216000 (60×60×60) SOLID186 cubic 20-node elements in simulations. All the nodes of the beam elements of the reinforcement fibre network are enforced to have the same displacements of their closest nodes of the matrix elements by a two-step searching & coupling technique [22]. General periodic boundary conditions are applied at the boundaries of the cubic RVE. In addition, it is worth to mention that the rotational degrees of freedom of the fibre nodes coupled on the facets, edges or vertices of the RVEs are constrained in the same periodic way as the translational degrees to achieve perfect PBCs.

In order to obtain the Young's Modulus of the composite $E_c$, a small uniaxial tensile strain of 0.1% is applied to the RVE model. To simplify the post-processing of the results, the Young's modulus of the matrix material is assumed to be $E_m = 1$. For the reinforcement fibre network made of a solid material, its Young's modulus $E_f$ is converted to an equivalent Young's modulus $\overline{E_f} = E_f - E_m$ in finite element simulations. As the periodic reinforcement fibre networks in the RVEs of the IPCs are randomly generated, all the simulation results presented in the figures below are the averaged values of 10 similar random RVEs.

## 3. Results

As the elastic properties of random 3D Voronoi fibre networks (i.e. Voronoi open cell foams) are isotropic [10,20], the elastic properties of IPCs made of an isotropic matrix and an isotropic reinforcement Voronoi fibre network are also isotropic in elastic properties. In this paper, all the presented simulation results (i.e. data points in the figures below) of the IPCs are the mean results obtained from 10 similar random RVE models with the same combinations of different parameters. The relationship between the Young's modulus of the IPCs and the fibre volume fraction of the reinforcement fibre network depends on different factors such as the number of complete cells, the degree of cell regularity and the defects (i.e. missing fibres) of the reinforcement fibre network in the RVE model, and the combination of the elastic properties of the constituent materials. The effects of these factors are respectively presented in the following subsections.

### 3.1. Effects of the number of complete cells in the reinforcement fibre network

Figure 5 shows the effects of the cell number of reinforcement fibre network on the Young's moduli of the IPCs with multiple reinforcement fibre volume fractions, where the elastic properties of the constituent materials are fixed at $E_f/E_m = 10$ and $\nu_f = \nu_m = 0.3$, and the degree of the reinforcement fibre network regularity is fixed at R=0.5. As can be seen from Figure 5, the larger the number of the complete cells in the reinforcement fibre network of the RVE model, the smaller the error bar (i.e. standard deviation of the results obtained from 10 similar models) of the Young's modulus of the IPCs. When N is sufficiently large (i.e. 54 or larger), the Young's moduli of the IPCs with the fixed fibre volume fraction are independent of the number of complete cells of the fibre network in the RVE model. Thus, in order to minimise the computational cost while simultaneously maintaining the accuracy, all the results presented below are obtained from RVE models of IPCs with a reinforcement fibre network whose number of complete cells is fixed at 54.

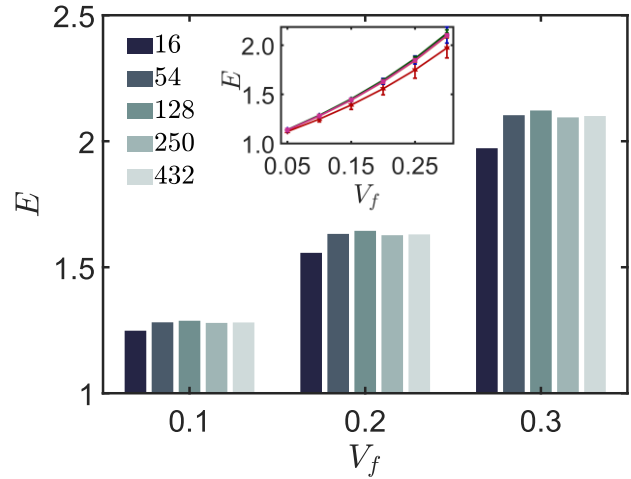

Figure 5. Effects of the number of complete cells in the reinforcement fibre network on the relationship between the Young's modulus of IPCs and the fibre volume fraction.

## 3.2. Effect of the degree of cell regularity

It has been found that the degree of cell regularity could significantly affect the elastic properties of random Voronoi open cell foams [9,10,20]. As the IPCs discussed in this paper are reinforced by such a Voronoi open cell foam (fibre network), the degree of cell regularity, $R$, can significantly affect the elastic properties of the IPCs. For IPCs reinforced by fibre networks with a fixed number of 54 complete cells (i.e. $N = 54$), the effects of the degree of cell regularity on the relationship between the Young's modulus of the IPCs and the fibre volume fraction are shown in Fig. 6, where the reinforcement fibre network is fully random when R=0.0, and a perfect regular BCC structure when R=1.0. Figure 6 indicates that the larger the degree of the cell regularity, the larger the Young's modulus of the IPCs.

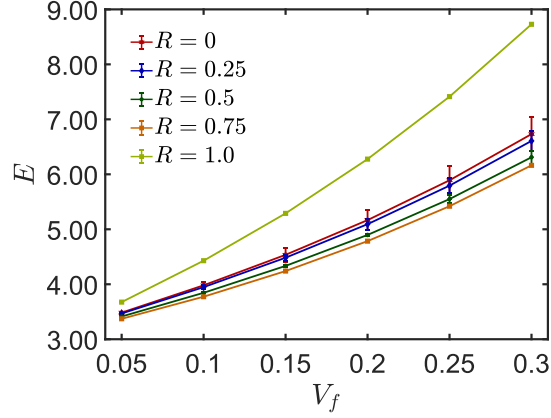

Figure 6. Effects of the degree of fibre network cell regularity on the relationship between the Young's moduli of the IPCs and the fibre volume fractions.

### 3.3. Effect of missing fibres

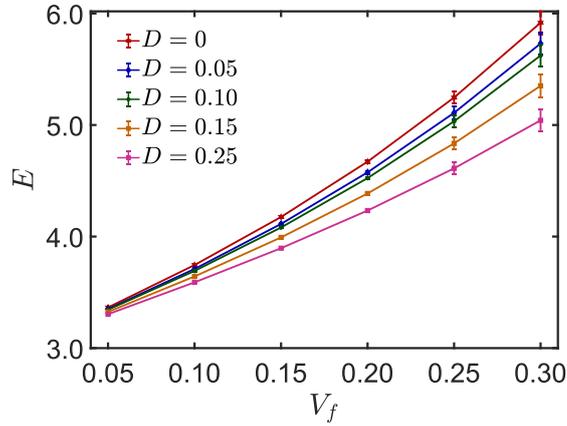

Figure 7. Effects of missing fibres on the relationship between the Young's moduli of the IPCs and the fibre volume fractions.

Missing fibres are a common type of defects in cellular materials, which have been found to significantly affect the mechanical properties of the cellular materials [23]. We randomly remove a certain percentage, $D$, of the total numbers of struts (fibres) from the reinforcement fibre network (open cell foam) while keeping the fibre volume fractions in the RVE model unchanged. To demonstrate the effects of missing fibres on the Young's modulus of IPCs, different percentages of missing fibres $D$ = 3%, 5%, 10%, 20% and 25% are randomly removed from the same reinforcement fibre network with a degree of regularity $R$=0.5. Figure 7 shows that the larger the percentage of the missing fibres, the larger the reduction of the Young's modulus of the IPCs. With the increase of the fibre volume fraction, the reduction amplitude of the Young's moduli of IPCs obviously increases, however, the relative reduction in amplitude of the Young's moduli remain almost unchanged.

by random fibre removing while keeping the volume fractions unchanged. It can be seen from Figure 7 that the defects of the reinforcement could lower the elastic performance of the composite (4.32% at $V_f = 0.1$ and 17.3% at $V_f = 0.3$) even with the same volume fraction of the reinforcement.

### 3.4. Effects of the elastic properties of the constituent materials

Figure 8 shows the effects of the elastic properties of the constituent materials, i.e. $E_f/E_m$, $\nu_f$ and $\nu_m$, on the relationship between the reinforcement fibre volume and the Young's modulus of IPCs with $E_m = 1$, $N=54$, $D=0$ and $R=0.5$. The Young's moduli of the IPCs, $E_c$, are calculated by averaging the $E_{cx}$, $E_{cy}$ and $E_{cz}$, and the normalized Young's moduli, $E_n$, are obtained as $E_c/E_{cVoigt}$.

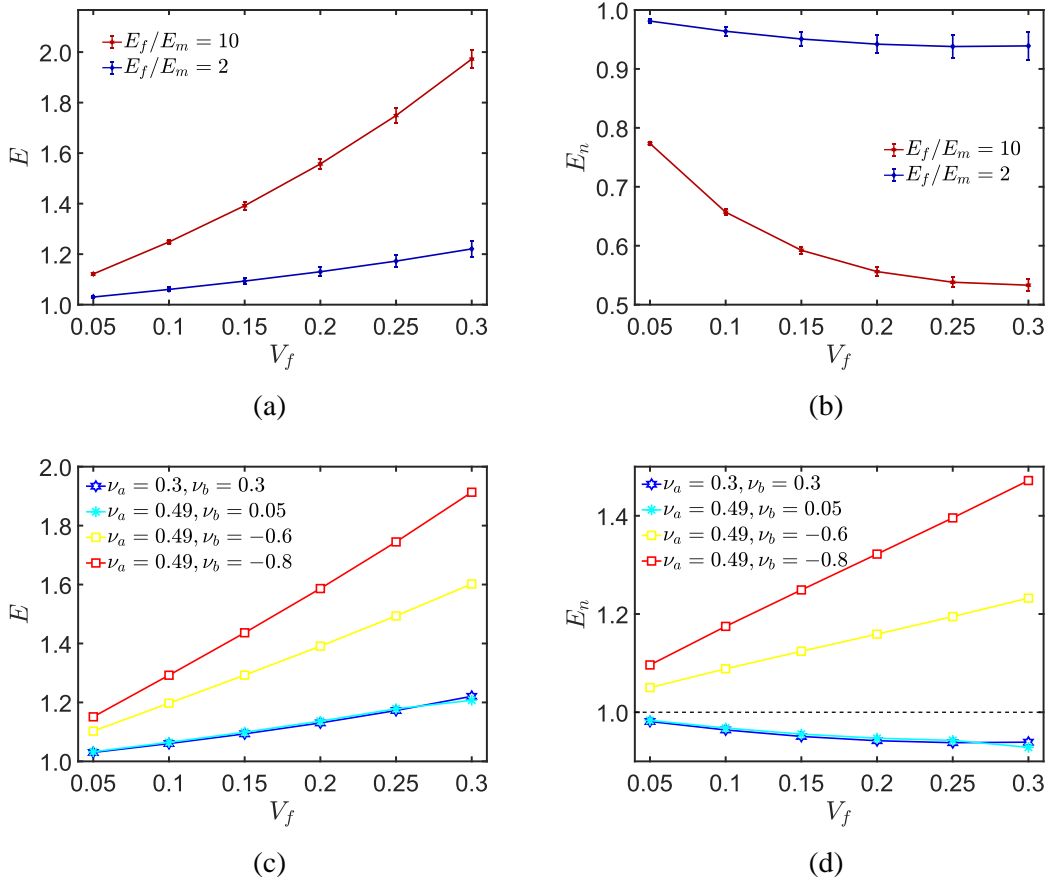

(a) (b) (c) (d)

Figure 8. Effects of the elastic properties of the constituent materials on the relationship between the fibre volume fractions and the normalised Young's modulus of IPCs with $N = 54$, $D = 0$ and $R=0.5$. (a) effects of $E_f/E_m$ on $E_c$ when $v_m = v_f = 0.3$, (b) effects of $E_f/E_m$ on $E_n$ when $v_m = v_f = 0.3$, (c) effects of $v_f$ and $v_m$ on $E_c$ when $E_f/E_m = 2$, (d) effects of $v_f$ and $v_m$ on $E_n$ when $E_f/E_m = 2$.

To explore the effects of the Young's moduli of the constituent materials, their Poisson's ratios are fixed at $v_f = v_m = 0.3$. Figure 8 (a) indicates that the Young's modulus of the IPCs increases with the increase of the fibre volume fraction. The larger the values of the fibre volume fraction and $E_f/E_m$, the larger the Young's modulus of the IPCs and the error bar. In contrast, larger values of the fibre volume fraction and $E_f/E_m$ result in smaller normalized Young's modulus $E_n = E_c/E_{cVoigt}$ of the IPCs as illustrated in Figure 8 (b). The normalised Young's moduli of the IPCs when $E_f/E_m = 2$ are much larger than those when $E_f/E_m = 10$, which suggests that the normalised Young's modulus $E_n = E_c/E_{cVoigt}$ becomes the maximum when the ratio of $E_f/E_m$ approaches 1.

To investigate the effects of the Poisson's ratios of the constituent materials, the ratio of the Young's moduli of the constituent materials is fixed at $E_f/E_m = 2$. Figures 8 (c) and (d) show how the different combinations of the Poisson's ratios of the constituent materials affect the relationships between fibre volume fraction and the Young's modulus $E_c$ as well as the normalised Young's modulus of the IPCs. As can be seen, the larger the difference between the Poisson's ratios of the constituent materials, the larger the Young's modulus and the normalised the Young's modulus $E_n = E_c/E_{cVoigt}$ of the IPCs. When the Poisson's ratio of the fibre material is positive and that of the matrix is negative, the normalized Young's moduli of the IPCs are large than 1, which means that the Young's modulus of the IPCs reinforced by a random 3D Voronoi fibre network can larger than the Voigt limit. These results are also consistent with the results in our previous work [8,24].

## 4. Discussion

One of the main advantages of the IPCs reinforced by a random Voronoi fibre network is that they are very easy to produce, and yet they still have better Young's modulus than those of the conventional composites. In the following, we will compare the Young's modulus of the IPCs reinforced by a random Voronoi fibre network with those of the conventional particle and fibre reinforced composites, and those of the IPCs reinforced by different types of regular cellular or fibre-network structures.

**4.1. Comparison with the conventional particle and fibre reinforced composites**

For the random particle and short fibre reinforced composites [25], the particle/fibre aspect ratios are defined as the ratio of their largest dimension to their smallest dimension. Thus $\xi = 1$ means spherical particle reinforcement and $\xi = 5$ means short fibres reinforcement. As the reinforcements are not fibre, the subscript $r$ is used here instead of $f$ in the previous sections. In literature [25], the reinforcement volume fraction is $V_r = 0.2$, and the elastic properties of the constituent material are $E_r = 70\text{GPa}$, $v_r = 0.2$ and $E_m = 3\text{GPa}$, $v_r = 0.35$, respectively. Figure 9 (a) and (b) are the periodic RVE models of this work [25], and Figure 9 (c) shows the periodic RVE model of the composite reinforced by completely randomly oriented fibres in literature [26]. The constituent materials of fibre reinforced composite in [26] are AS4 carbon fibre and 3501-6 Epoxy matrix and the elastic properties are $E_r = 225\text{GPa}$, $v_r = 0.2$ and $E_m = 4.2\text{GPa}$, $v_r = 0.35$. The volume fractions were $V_r = 15.23\%$, 19.23% and 21.64% for RVE generation in Babu *et al.*'s work. The geometry of this work is shown in Figure 9 (c).

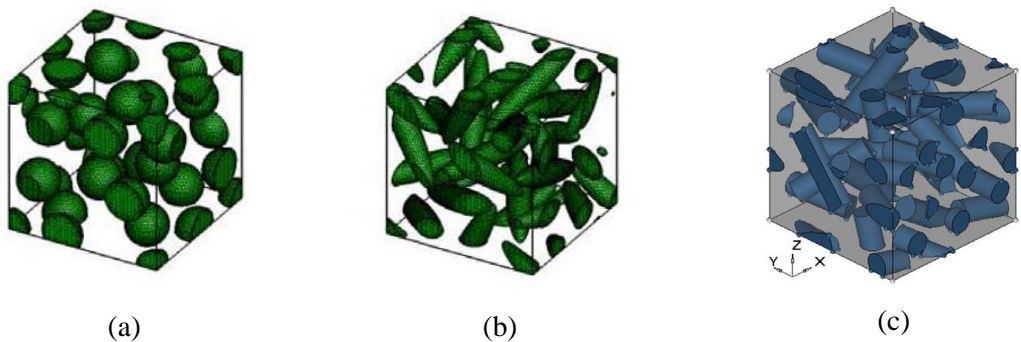

Figure 9. The RVE geometric models of isotropic composites reinforced by randomly distributed particles or fibres: (a) Ref [25], $\xi = 1$. (b) Ref [25], $\xi = 5$. (c) Ref [26], Case 4.

In order to compare our results of IPCs reinforced by a random Voronoi fibre network with those of the literature [25, 26], the same elastic properties and volume fractions of the constituent materials are applied to the RVE models of the IPCs, the degree of the cell regularity is fixed at $R = 0.5$ and the number of the complete Voronoi cells is fixed at $N = 54$. The comparisons between the Young's moduli of the composites in literature [25] and our results of IPCs are presented table 1 and 错误!未找到引用源。, respectively. As can be seen from Table 1, the Young's modulus $E_c$ of the IPC reinforced by a random Voronoi fibre network is 18.7% larger than those of the composites reinforced by either random spherical particles ($\xi = 1$) or by random short fibres ($\xi = 5$) in literature [25]. Besides, Figure 10 shows clearly that the Young's moduli $E_c$ of the IPCs reinforced by a random Voronoi fibre network in this work are over 80% larger than the simulation results of the composites reinforced by short fibres in literature [26]. Thus, the IPCs in this work are not only easy to produce, but more importantly have much better mechanical properties than the conventional particle and short fibre reinforced composites.

Table 1. Comparison between the Young's moduli of the composites in literature [25] and our result of IPC.

| Composite | $VF_r = 0.2$ | $E_c$ |
| --- | --- | --- |
| Reference [25], $\xi = 1$ | 0.2 | 6.6233 |
| Reference [25], $\xi = 5$ | 0.2 | 6.7700 |
| IPC in this work | 0.2 | 7.8970 |

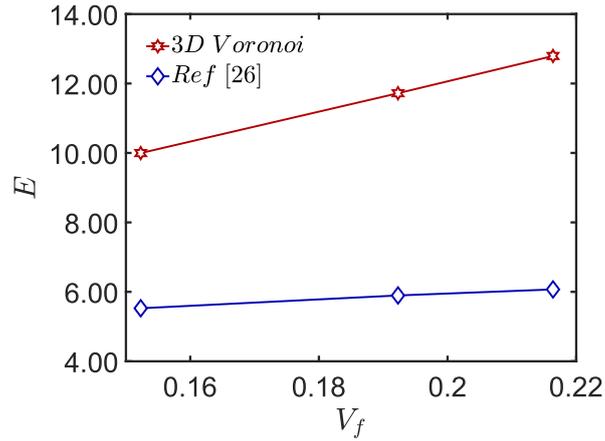

Figure 10. Comparison between the Young's moduli of the composites reinforced by randomly distributed short fibres in case 4 of literature [26] and those of the IPCs reinforced by a random Voronoi fibre network.

## 4.2. Comparison with other types of IPCs

Furthermore, it is necessary to compare different lattice structure reinforced composites in this thesis to determine which one have the best elastic performance under different constituent materials. To compare the Young's moduli of the composites, the regular lattice structured IPCs and auxetic lattice structured IPCs with fibre concavity angle $\alpha = 18°$ and the 3D Voronoi composite created with coefficient of regularity $R = 0.5$ and Voronoi cell number $N = 54$ in this work are listed for comparison below in Figure 11.

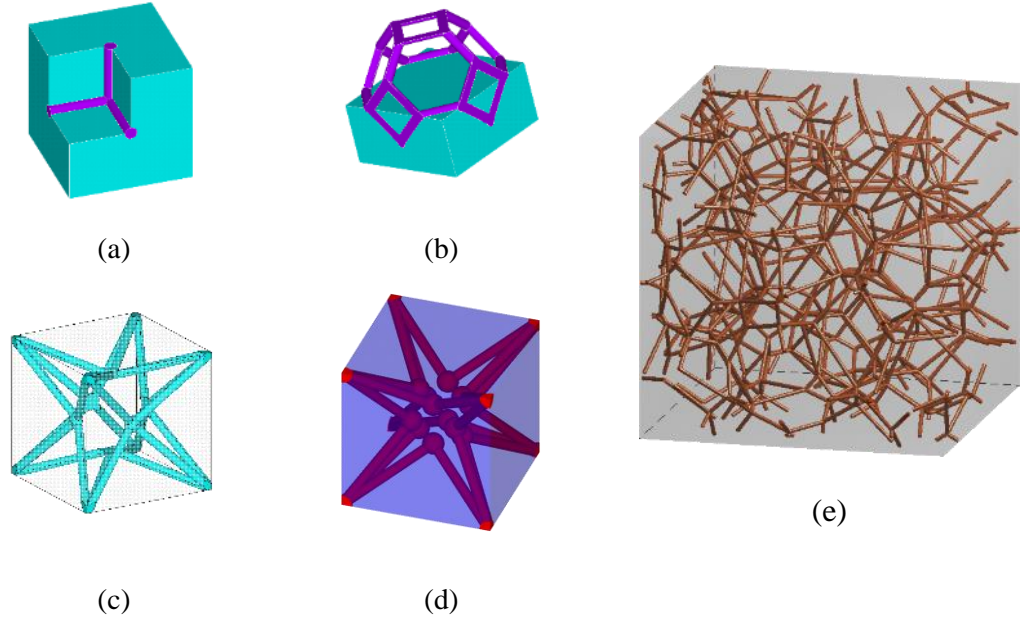

Figure 11. RVEs of lattice structure reinforced interpenetrating composites listed for comparison. (a) cubic lattice reinforced IPC; (b) tetradecahedron cell reinforced IPC; (c) auxetic lattice reinforced IPC structure I; (d) auxetic lattice reinforced IPC structure II; (e) 3D Voronoi fibre network reinforced IPC.

Four sets of parameters for the constituent materials as listed in Table 2: the first two sets represent strong fibre and polymer matrix combination such as carbon/epoxy or aluminium/epoxy; the last two sets represent metal-metal or polymer-polymer composites which the fibres and the matrix have closer mechanical properties.

Table 2. Three sets of constituent material parameters.

|  | $E_f$ | $E_m$ | $v_f$ | $v_m$ |
| --- | --- | --- | --- | --- |
| Set 1 | 100 | 1 | 0.1 | 0.3 |
| Set 2 | 100 | 1 | 0.1 | -0.5 |
| Set 3 | 5 | 1 | 0.1 | 0.3 |
| Set 4 | 5 | 1 | 0.1 | -0.5 |

The Young's moduli of different types of IPCs are plotted in Figure 12.

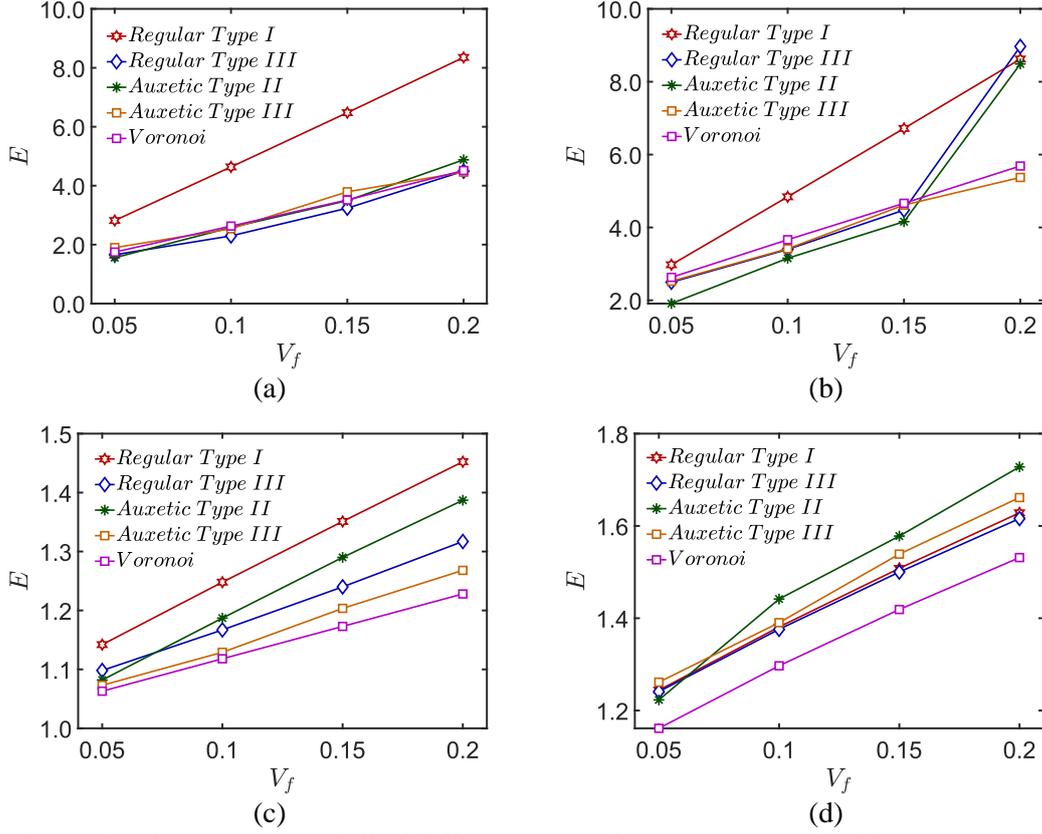

Figure 12. The Young's moduli of different types of IPCs constructed with same constituent materials. (a) $E_f/E_m = 100$, $\nu_f = 0.1$, $\nu_m = 0.3$ (b) $E_f/E_m = 100$, $\nu_f = 0.1$, $\nu_m = -0.5$ (c) $E_f/E_m = 5$, $\nu_f = 0.1$, $\nu_m = 0.3$ (d) $E_f/E_m = 5$, $\nu_f = 0.1$, $\nu_m = -0.5$.

An overview of Figure 12 (a), (b) and (c) shows that the best structure in terms of stiffness is apparently the cubic lattice reinforced IPC when $\nu_f$ and $\nu_m$ is positive. When $\nu_m$ is negative, the stiffness of cubic lattice reinforced composite is the largest among those of different IPC structures at 5%, 10% and 15% fibre volume fractions. However, the stiffness of tetradecahedron cell reinforced IPC and that of auxetic IPC structure II come close to the stiffness of cubic lattice reinforced IPC at 20% volume fractions. The only condition to consider other type of IPCs to be the best is when $E_f/E_m = 5$, $\nu_f = 0.1$, $\nu_m = -0.5$. Interestingly, the auxetic IPC structure II and tetradecahedron cell reinforced IPC show the similar stiffness behaviour, while the stiffness as a function of the fibre volume fraction of auxetic interpenetrating composite type III and that of the 3D Voronoi fibre network reinforced composite type II ($R = 0.5$, $N = 54$) are similar. This can be found in all the subfigures of Figure 12, especially in (b) and (d).

4.3. Voronoi fibre network reinforced model to simulation aerogel

Aerogel is a class of low-density, sol-gel-derived, nanoporous cellular materials [27] with ultralow thermal conductivities [28], extremely low densities [29] and other record-breaking physical properties [27]. The material of the cellular structure of the aerogels could be silica [30–34], carbon [35–37] and a variety of other materials [38–40]. The Voronoi fibre network reinforced model purposed in this paper could be an approach to investigate the mechanical behaviours of the aerogels. The biopolymer-based aerogels with different κ-carrageenan [19] wt% are simulated by our model with the data proposed in different references and the results below in shows that the elastic response of aerogel could be well-captured by our models.

Table 3. biopolymer-based aerogels with different κ-carrageenan wt%

| Skeletal density | 1.741 | 1.717 | 1.7184 |
| --- | --- | --- | --- |
| $V_f$ | 0.034 | 0.053 | 0.080 |
| $E_{ref}$ (MPa) | 5.097 | 9.756 | 18.676 |
| $E_{simulation}$ (MPa) | 5.840 | 9.903 | 16.280 |

## 5. Conclusion

The 3D Voronoi fibre network is generated via Voronoi tessellation and is used as the reinforcement of the composite. Compared to regular lattice structures, the 3D Voronoi network could give a better description of some fabricated interpenetrating composites. The periodicity of the random fibre network is ensured by filling multiple tessellation space with groups of Voronoi points. and then divided them into fully periodic RVEs. Furthermore, the coefficient of regularity $R$ and Voronoi points/cells contained in an RVEs $N$ is defined when constructing the Voronoi fibre network.

The 3D Voronoi fibre reinforced composite is nearly isotropic. The stiffness of the composite shows a linear relation with the fibre volume fraction. The Young's modulus of the composite $E_c$ increase as $E_f$ and $E_m$ increase. However, the study of the normalized Young's moduli $E_n$ indicates that $E_n$ increases when $E_f/E_m$ approaches 1 and $|v_f - v_m|$ get the largest possible value. However, as the volume fraction is limited up to 20%, no peak of $E_n$ is found in the range of 1% to 20% fibre volume fraction and $E_n$ increases with the increase of fibre volume fraction.

The comparison of existing journal results of random fibre reinforced composites shows that the 3D Voronoi composite performs better than the discrete particle and short fibre reinforced composites in terms of elastic moduli. To determine the best structure in different given conditions, the structures investigated is compared in terms of the stiffness. It is found that regular lattice structure is the most effective structure in build high stiffness composites in small fibre volume fractions. The auxetic IPC type II and regular lattice reinforced IPC type III shows the similar elastic behaviour as the fibre volume fraction changes, while the stiffness as a function of the fibre volume fraction of auxetic interpenetrating composite type III and that of the 3D Voronoi fibre network reinforced composite type II ($R = 0.5, N = 54$) are similar.

**Appendix I**

Mesh sensitivity of the element coupling in this work

The coupling quality of the nodes of the fibres and matrix is largely affected by the element size. When meshed with large elements, it is more possible that the distance between closest $N_m$ for $N_f$ is too far to obtain proper results. However, the element size used in mesh cannot be too small as the solving time of the model is unacceptable for a very large meshed model. RVE length $L = 30$ is used in all these Voronoi fibre networks. To give an overview of the effect of element size used in the mesh of fibres and matrix on the mechanical properties of the composite, different combinations of $es_f$ and $es_m$ are considered in this section. Instead of different sizes of element used in the matrix and fibres, the ratio between the fibre and matrix is what matters most. The ratio between the element sizes of matrix and fibre with the length of the cube affects the result as well. Different element sizes of the fibres with beam elements are given in Table 4, while the element size of the matrix is set as 1. The coefficient of regularity is chosen to be $cor = 0.5$. The number of Voronoi points and cells presenting the number of fibres and volume fraction used in the mesh sensitivity check are $n = 4 \times 4 \times 4 = 64$ and $f_f = 10\%$, respectively. The other parameter such as mechanical properties of the constituent materials are specified as $E_m = 1$, $E_f = 100$, $\nu_m = 0.3$ and $\nu_f = 0.1$.

Table 4. Element size and Young's moduli obtained in Mesh Group 1

| Mesh Number | 1 | 2 | 3 | 4 | 5 | 6 | 7 | 8 | 9 | 10 |
|---|---|---|---|---|---|---|---|---|---|---|
| $es_f$ | 0.0625 | 0.125 | 0.25 | 0.5 | 1 | 2 | 4 | 8 | 16 | 32 |
| $es_m$ | 1 | 1 | 1 | 1 | 1 | 1 | 1 | 1 | 1 | 1 |
| $E$ | 59.79 | 30.22 | 15.44 | 8.32 | 4.51 | 2.58 | 2.01 | 1.95 | 1.95 | 1.95 |

With the element size combination of Mesh Group 1 shown in Table 4, the Young's moduli of the same RVE in $x$ direction $E_x$ are also listed. A more direct illustration is given in Figure A1(a).

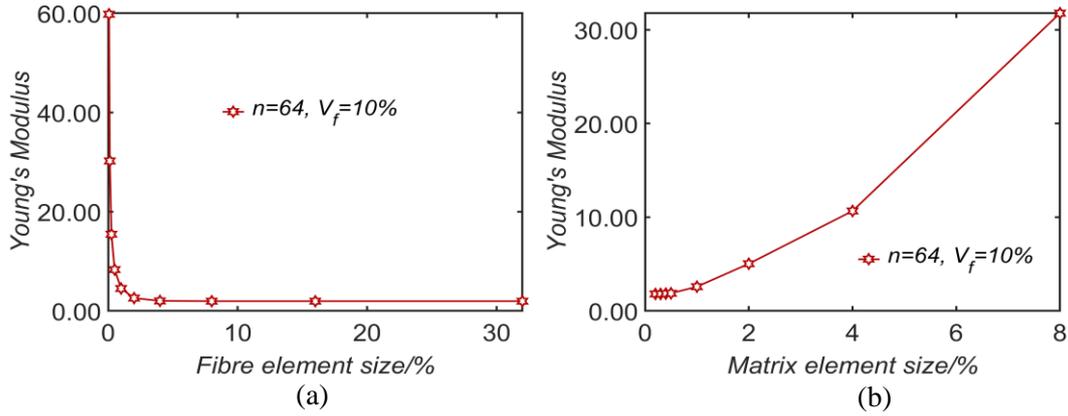

Figure A1. Mesh sensitivity of ASC technique for 3D Voronoi fibre network reinforced IPCs (a) Young's moduli of the composite as a function of fibre element size. (b) Young's moduli of the composite as a function of matrix element size.

It can be seen that if fibre element size is relatively small while the matrix element size is relatively large, the Young's Moduli of the composite obtained can be extraordinarily large. It is almost impossible for a random fibre reinforced composite with fibre volume fraction $V_f = 10\%$ to reach that a large number, almost 5 times larger than its Voigt limit $(E_c)_{Voigt} = 10.9$. This can be caused by the big difference between fibre element size and matrix element size, which lead to the constraint of different nodes on the fibres to the same nodes on the matrix. When the fibre element size is as small as around 0.1 while the matrix size is 1 as the mesh 1 and mesh 2 shown in Table 4, that means 10 nodes of the fibre will be constrained on the same node of the matrix. That is highly over constrained than that to represent a normal fibre-matrix interface, thus lead to the extreme overestimation of the fibre stiffness in Figure A1(a). When the fibre mesh size is large enough (larger than 5 in Figure A1(a)), even larger than the length of most of the fibres, the fibres are all meshed with a single beam element. As the mesh condition is the same for one element per fibre, the results are steady. Another mesh group is given in Table 5 as the element size of the fibre remains unchanged while the element size of the matrix varies from 1/10 to 4 times of the fibre element size. The results are provided in Figure A1(b).

Table 5. Element size and Young's moduli obtained in Mesh Group 2

| Mesh Number | 1 | 2 | 3 | 4 | 5 | 6 | 7 | 8 |
|---|---|---|---|---|---|---|---|---|
| $es_f$ | 2 | 2 | 2 | 2 | 2 | 2 | 2 | 2 |
| $es_m$ | 0.2 | 0.3 | 0.4 | 0.5 | 1 | 2 | 4 | 8 |
| Young's moduli of the composite | 1.805 | 1.809 | 1.838 | 1.878 | 2.583 | 5.030 | 10.661 | 31.774 |

According to Table 4 and Table 5 above, rules of element size can be summarized. Firstly, a proper element size combination of the two constituent materials must be with a relatively large $es_f/es_m$ to make the ASC coupling nodes as close as possible. This can avoid overestimation of the elastic properties of the composite. Secondly, the $es_f$ need to be small enough to give a good description of the fibres. Finally, as the $es_f$ need to be small enough and $es_f/es_m$ need to be large, the corresponding $es_m$ need to be a smaller number as well. Thus, a balance between the calculation efficiency and the accuracy need to be considered.

Following these three rules, three different combinations of $es_f$ and $es_m$ are selected in Table 6. ASC technique with these $es_f$ and $es_m$ combinations are compared with pure solid element models of structures Type I and Type III used in *Ref*.

Table 6. Three different combinations of $es_f$ and $es_m$

| Combination number / Element size | Set 1 | Set 2 | Set 3 |
|---|---|---|---|
| $es_f$ | 2 | 1 | 1 |
| $es_m$ | 1 | 0.5 | 0.2 |

Figure A2(a) and (b) illustrates the RVEs of Type I and Type III built by ASC technique with beam elements for the fibres and solid elements for the matrix.

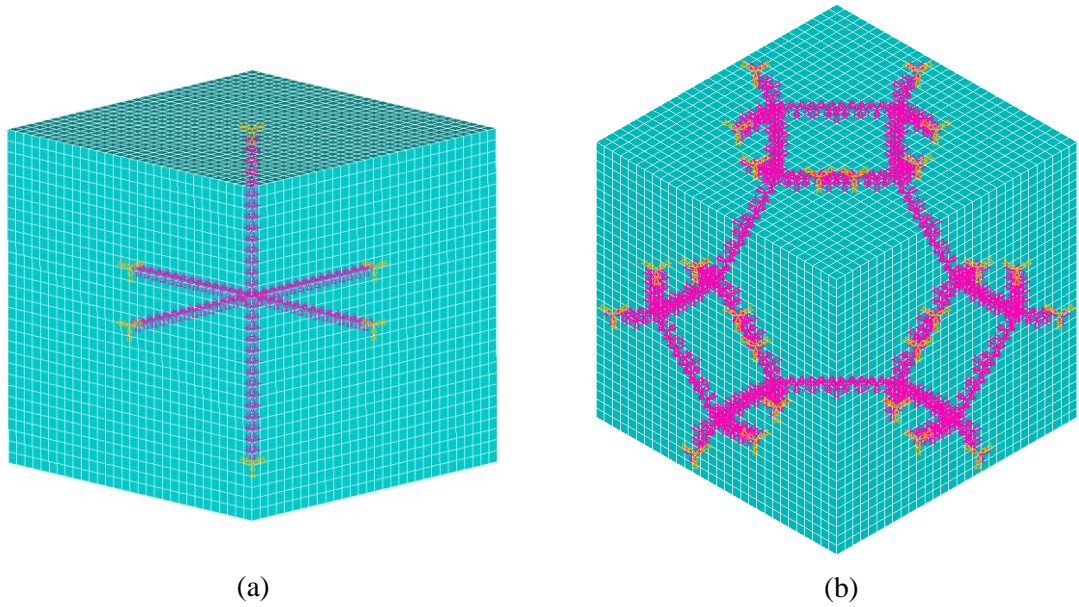

|(a)|(b)|

Figure A2. Illustrations of RVEs of Type I (a) and Type III (b) structures built by ASC technique with beam elements for the fibres and solid elements for the matrix.

It is worthwhile to mention that Type III structure cannot be modelled with the same RVE selection as that in *Ref* because solid element fibres can be intersected in axial direction while the beam element fibres cannot.

Figure A3 gives a comparison of the composite elastic moduli predicted by ASC coupled models and full solid models.

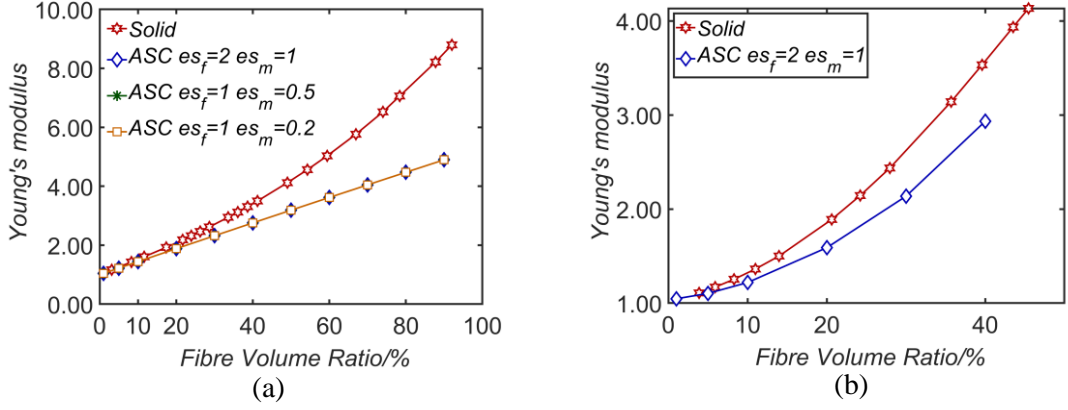

Figure A3. Comparisons of $E_c$ predicted by ASC coupled models and full solid models. (a) regular structured model Type I. (b) regular structured model Type III.

In Figure A3(a), the prediction of $E_c$ under different $es_f$ and $es_m$ settings are almost exactly the same. However, in Figure A3(a), the difference between the full solid model and ASC model becomes larger as the fibre volume fraction increases. As we only consider volume fractions between 1% and 20%, the accuracy of ASC technique is acceptable. The largest difference between the results of solid model and ASC model is less than 10% in Figure A3(a) and is less than 13% in Figure A3(b), when $es_f = 2$ and $es_m = 1$. The results are slight better when $es_f = 1$ and $es_m = 0.5$, but the calculations time is much longer. In summary, a combination of element sizes $es_f = 2$ and $es_m = 1$ is proper for the simulations.

2015;1:256–75.

[33] He S, Chen X. Flexible silica aerogel based on methyltrimethoxysilane with improved mechanical property. J Non Cryst Solids 2017;463:6–11. .

[34] Gurav JL, Jung I, Park H, Kang ES, Nadargi DY. Silica Aerogel : Synthesis and Applications 2010;2010.

[35] Worsley MA, Pauzauskie PJ, Kucheyev SO, Zaug JM, Hamza A V, Jr JHS, et al. Properties of single-walled carbon nanotube-based aerogels as a function of nanotube loading 2009;57:5131–6.

[36] Worsley MA, Pauzauskie PJ, Olson TY, Biener J, Satcher JH, Baumann TF. Synthesis of Graphene Aerogel with High Electrical Conductivity 2010:14067–9.

[37] Santos-gómez L, García JR, Montes-morán MA, Menéndez JA, García-granda S, Arenillas A. Ultralight-Weight Graphene Aerogels with Extremely High Electrical Conductivity 2021;2103407.

[38] Zhao S, Malfait WJ, Guerrero-Alburquerque N, Koebel MM, Nyström G. Biopolymer Aerogels and Foams: Chemistry, Properties, and Applications. Angew Chemie - Int Ed 2018;57:7580–608.

[39] Rege A. Micro-mechanical modelling of cellulose aerogels from molten salt hydrates 2016;12.

[40] Cifuentes A, Itskov M, Santos-rosales V, Alvarez-rivera G, Hillga M, Garc CA, et al. Stability Studies of Starch Aerogel Formulations for Biomedical Applications¨ 2020.

[41] M. W. D. Van-Der-Burg, Shulmeister V, Van-Der-Geissen E. On the Linear Elastic Properties of Regular and Random Open-Cell Foam Models. J Cell Plast 1997.

[42] Deogekar S, Picu RC. On the strength of random fiber networks. J Mech Phys Solids 2018;116:1–16. https://doi.org/10.1016/j.jmps.2018.03.026.

[43] Negi V, Picu RC. Mechanical behavior of cross-linked random fiber networks with inter-fiber adhesion. J Mech Phys Solids 2019;122:418–34.
28